\documentclass[aps,twocolumn,superscriptaddress,showpacs]{revtex4}
\usepackage{amssymb}
\usepackage{graphicx}

\input{tcilatex}

\begin{document}

\title{Structural Trends Interpretation of the Metal-to-Semiconductor Transition in
Deformed Carbon Nanotubes}
\author{Jun-Qiang Lu}
\affiliation{Center for Advanced Study and Department of Physics, Tsinghua University,
Beijing 100084, China}
\affiliation{Department of Mechanical and Industrial Engineering, University of Illinois, 
Urbana, Illinois 61801}
\author{Jian Wu}
\affiliation{Center for Advanced Study and Department of Physics, Tsinghua University,
Beijing 100084, China}
\author{Wenhui Duan}
\affiliation{Center for Advanced Study and Department of Physics, Tsinghua University,
Beijing 100084, China}
\author{Bing-Lin Gu}
\affiliation{Center for Advanced Study and Department of Physics, Tsinghua University,
Beijing 100084, China}
\author{H. T. Johnson}
\affiliation{Department of Mechanical and Industrial Engineering, University of Illinois, 
Urbana, Illinois 61801}
\date{\today }

\begin{abstract}
Two mechanisms that drive metal-to-semiconductor transitions in
single-walled carbon nanotubes are theoretically analyzed through a simple
tight-binding model.  By considering simple structural trends, the results demonstrate that metal-to-semiconductor
transitions can be induced more readily in metallic zigzag nanotubes than in
armchair nanotubes.  Furthermore, it is shown that both mechanisms have the effect of making the two originally equivalent sublattices 
\textit{physically }distinguishable.
\end{abstract}

\maketitle





Carbon nanotubes are promising candidates for the next generation of
nanometer-scale electronic devices due to their unique electronic properties.%
\cite{SI,JWM}  As the electronic properties of an ideal carbon nanotube are 
uniform throughout the structure, the key problem for electronic device development 
is to control or tailor the electronic properties of a given nanotube.

Experimentally, atomic force microscope tips are used to manipulate 
single-walled carbon nanotubes (SWNTs).\cite{TWT}  The results show that
mechanical deformation of SWNTs may lead to a metal-to-semiconductor
transition (MST) in their electronic properties.  This experiment suggests 
a means for SWNT-based electronic devices:  a metallic SWNT can act as a good
conducting lead, and when mechanical deformation leads to semiconducting
areas along the tube a nanometer-scale electronic device is produced.  Thus,
understanding MSTs in SWNTs is a necessary step in realizing nanotube
based electronic devices.

Theoretically, many studies consider the MST in SWNTs. Park 
\textit{et al. }point out that the breaking of the mirror symmetry 
leads to a MST in armchair SWNTs.\cite{CJP} Lammert \textit{et al. }compare the squashed armchair SWNTs to graphene bilayers and explain the
MST in squashed armchair SWNTs through the interaction between graphene
bilayers.\cite{PEL} For metallic zigzag SWNTs, all the studies agree that
the MST is driven by the curvature effect.\cite{CJP,PEL}

\begin{figure}[b]
\begin{center}
\includegraphics[width=3.0in]{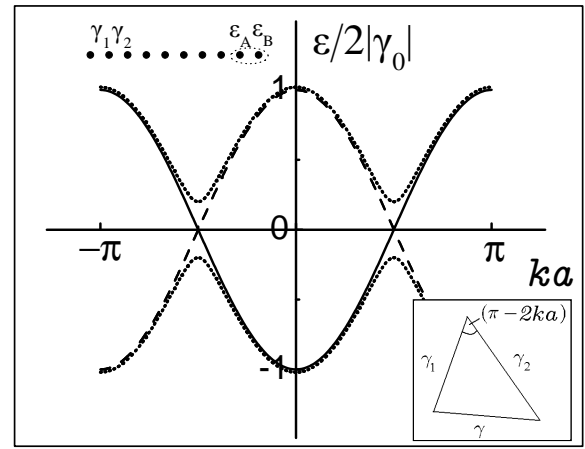}
\end{center}
\caption{Solid line: the energy dispersion relation
of the one-dimensional lattice system.  Dotted line: the energy dispersion
relation of the system shown in the left-top corner.  Lower right inset:
the relationships between $\protect\gamma $, $\protect\gamma _{1}$, and $%
\protect\gamma _{2}$.}
\end{figure}

In the authors' previous work,\cite{JQL} the MST in squashed armchair
SWNTs is explored.  It is shown that when a structural perturbation is made the two original
equivalent sublattices in the armchair SWNTs become distinguishable, 
and an energy gap opens, leading to a MST.  It is also pointed out that the \textit{physical }distinction of the two sublattices in armchair SWNTs could be achieved by a
combined effect of mirror symmetry breaking and bond formation between the
flattened faces of the squashed tubes.  An unresolved issue, then, is whether the mechanism can drive a MST in
metallic zigzag SWNTs.  Experimental measurements and theoretical
simulations both show that a MST can be achieved more easily in metallic zigzag
nanotubes than armchair nanotubes,\cite{CJP,PEL,EDM,JC} but a consistent 
explanation of the mechanism driving the MST is missing.  In this paper, a unified 
explanation is presented for the MST in SWNTs based on a simple tight-binding
model.  Simple structural trends based on the tight-binding model are used to explain the MST.  
First, two different mechanisms that can drive a
MST in SWNTs are presented, including i) a difference in the onsite energies of the two
sublattices and ii) a difference in the nearest-neighbor hopping
integral along the axis and along the circumference.  It is shown that only the first mechanism
can lead to the MST in armchair SWNTs, while both mechanisms can
drive a MST in metallic zigzag SWNTs.  It is also shown that it is the second
mechanism that drives the MST in squashed metallic zigzag SWNTs.  
Last, it is pointed out that the effects of the two mechanisms are unified: both make the
two original equivalent sublattices \textit{physically }distinguishable. 
In general, the distinction of the two sublattices is the mechanism that
drives a MST in any metallic SWNT.  This is also the case for a graphene monolayer, 
but, incidentally, \textit{not }the case for a graphene bilayer.  This illustrates a 
key difference between a SWNT and a graphene bilayer.

First, a tight-binding model of a one-dimensional
lattice system is considered.  The energy dispersion relation of the system, 
illustrated as the solid line in Figure 1, is a simple
cosine function given by 
\[
\varepsilon -\varepsilon _{0}=2\gamma _{0}\cos (ka), 
\]%
where, $\varepsilon _{0}$ is the onsite energy, $\gamma _{0}$ is the
nearest-neighbor hopping integral, $k$ is the one-dimensional wave vector,
and $a$ is the lattice constant. For the case in which the bands are half filled, the Fermi
energy of the system is $\varepsilon _{0}$, and the curve of the energy
dispersion relation continues at the Fermi energy, so the system is metallic.

Now, if the one-dimensional system includes two sublattices with
different onsite energies $\varepsilon _{A}$ and $\varepsilon _{B}$, as
shown in the left-top corner of Figure 1, then the energy dispersion relation of
the system can be expressed as 
\[
\varepsilon -\varepsilon _{0}=\pm \{\Delta ^{2}+(2\gamma _{0}\cos
ka)^{2}\}^{1/2},
\]%
where, $\varepsilon _{0}=(\varepsilon _{A}+\varepsilon _{B})/2,\Delta
=\left\vert \varepsilon _{A}-\varepsilon _{B}\right\vert /2$. This is plotted
in Figure 1 as the dotted line.  It can be seen that an energy gap (with
value $2\Delta $) is open near the Fermi energy $\varepsilon _{0}$, so the
system is semiconducting.  From the two cases in this model, it is apparent
that a difference in the onsite energies of the two sublattices leads
to a MST. 

Next, another mechanism leading to a MST in the
one-dimensional system is examined: different values of the nearest-neighbor
hopping integral.  It is assumed first that the two nearest hopping intergrals are
changed to $\gamma _{1}$ and $\gamma _{2}$, as shown in the left-top corner
of Figure 1, while the onsite energies are kept as $\varepsilon _{0}$. The
energy dispersion relation of the system is then: 
\[
\varepsilon -\varepsilon _{0}=\pm \gamma , 
\]%
\[
\gamma =\left( \gamma _{1}^{2}+\gamma _{2}^{2}+2\gamma _{1}\gamma _{2}\cos
2ka\right) ^{1/2}, 
\]%
where $\gamma $ and $\gamma _{1}$, $\gamma _{2}$ can be treated as the three
sides of a triangle, as shown in the right-bottom inset of Figure 1.
To get $\gamma =0$, the two conditions: $2ka=\pi $ and $\gamma _{1}=\gamma
_{2}$, must be satisfied. So, if the two hopping intergrals $\gamma _{1}$
and $\gamma _{2}$ become too different, $\gamma $ will not be zero no matter
what value $k$ is.  Then the energy dispersion relation of the system will not 
be continuous at the Fermi energy $\varepsilon _{0}$.  An energy gap (with value $%
2\left\vert \gamma _{1}-\gamma _{2}\right\vert $) will open near the Fermi
energy $\varepsilon _{0}$ so the system will be semiconducting.

\begin{figure}[b]
\begin{center}
\includegraphics[width=3.0in]{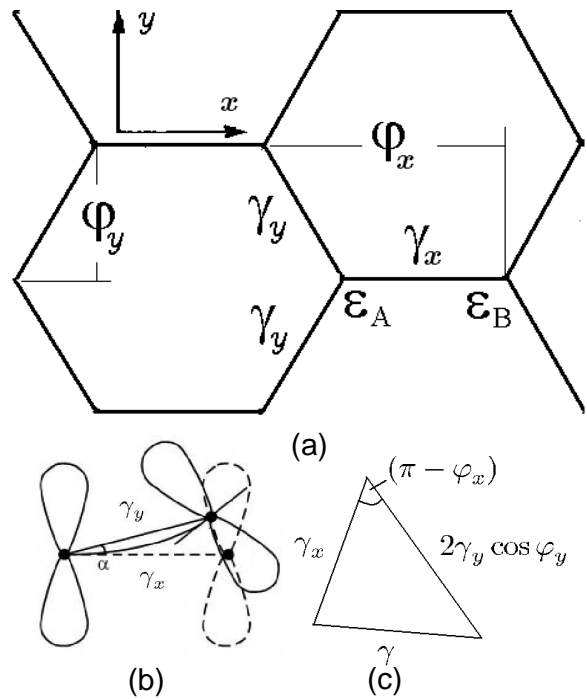}
\end{center}
\caption{(a) The graphene sheet-like structure with
different onsite energies and nearest-neighbor hopping integrals. (b) The
difference between $\protect\gamma _{x}$ and $\protect\gamma _{y}$ due to the
curvature effect. (c) The relationships between $\protect\gamma $, $%
\protect\gamma _{x}$, and $2\protect\gamma _{y}\cos \protect\varphi _{y}$.}
\end{figure}

From the above discussion, two mechanisms that can drive a MST
in the one-dimensional lattice system are identified: i) a difference in the onsite
energies of the two sublattices and ii) a difference in the
nearest-neighbor hopping integrals.  Furthermore, it can be said that the
effect of both mechanisms is to make the two original equivalent
sublattices \textit{physically }distinguishable.  Applying the same exercise to
SWNTs, it is found that only the first mechanism can lead
to a MST in armchair SWNTs, while a MST can be achieved by either mechanism in
metallic zigzag SWNTs.

It is well-known that a graphene sheet is metallic.  Here the processes of wrapping a graphene 
sheet into a nanotube and then squashing it are considered.  As shown in Figure 2b, the
curvature due to the wrapping will reduce the $pp\pi $\ overlap between the
nearest-neighbor carbon atoms to the original $\cos ^{2}\alpha $;\cite{MO} at the
same time, the $\sigma $ orbital and $\pi $ orbital between the
nearest-neighbor carbon atoms will not be normalized. These two changes lead
to different nearest-neighbor hopping integrals along the axis and
the circumference.  On the other hand, as shown previously,\cite{JQL}
structural perturbations can also introduce new interactions between the
atoms in one sublattice, to distinguish from those in the other sublattice, which
is equivalent to assigning different onsite energies to the two sublattices, as
is the case in Boron-Nitride nanotubes.\cite{ARU}  So, with the structural
perturbations, the system shown in Figure 2a will not be a graphene sheet any
longer, and it is assumed the two sublattices have different onsite energies $%
\varepsilon _{A}$ and $\varepsilon _{B}$, and nearest-neighbor hopping
integrals along the $x$ and $y$ directions given by $\gamma _{x}$ and $\gamma _{y}$%
, respectively.  Thus, the energy dispersion relation of the system can be
expressed as: 
\[
\varepsilon (k_{x},k_{y})-\varepsilon _{0}=\pm \{\Delta ^{2}+\gamma
^{2}\}^{1/2},
\]%
where, $k_{x}$ and $k_{y}$ are the wave vector components along the $x$ and $y$
directions, and 
\[
\varepsilon _{0}=(\varepsilon _{A}+\varepsilon _{B})/2,\Delta =\left\vert
\varepsilon _{A}-\varepsilon _{B}\right\vert /2,
\]%
\[
\gamma =\left( \gamma _{x}^{2}+4\gamma _{y}^{2}\cos ^{2}\varphi _{y}+4\gamma
_{x}\gamma _{y}\cos \varphi _{x}\cos \varphi _{y}\right) ^{1/2},
\]%
$\varphi _{x}$ and $\varphi _{y}$ are half of the phase increments between
the nearest-neighbor carbon atoms which belong to the same sublattice along
the $x$ and $y$ directions, as shown in Figure 2a, and $\varphi _{x}=\sqrt{%
3}k_{x}a/2$, $\varphi _{y}=k_{y}a/2$.

From the energy dispersion relation, to keep the system metallic, two
conditions must be satisfied: $\Delta =0$ and $\gamma =0$. Breaking of
either condition will lead to a MST.  Thus, if the two sublattices have
different onsite energies then $\Delta \neq 0$, and a MST can be achieved in not
only the armchair SWNTs, but also in all metallic SWNTs and even in the graphene
monolayer.

\begin{figure}[b]
\begin{center}
\includegraphics[width=2.4in]{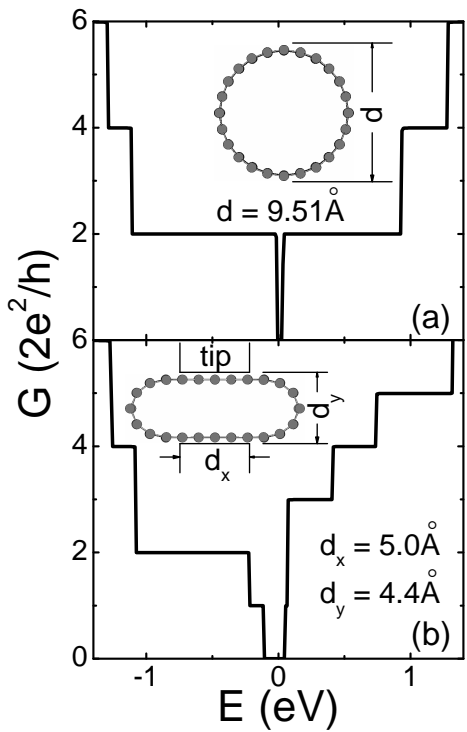}
\end{center}
\caption{Conductances of various nanotube structures,
which are shown as the insets.  $E$ is the energy of injected electrons, and
the Fermi energy of an ideal zigzag (12,0) nanotube is taken as zero.}
\end{figure}

The other condition is that $\gamma =0$.  In fact, as shown in
Figure 2c, $\gamma $ can be treated as one of the sides of a triangle,
where the other two sides are $\gamma _{x}$ and $2\gamma _{y}\cos \varphi
_{y}$.  So, the condition $\gamma =0$ can be divided into two conditions: $%
\varphi _{x}=\pi $ and $2\gamma _{y}\cos \varphi _{y}=\gamma _{x}$.  In the process 
by which a graphene sheet is wrapped into a SWNT it is found that $\varphi _{x}=\pi $ 
can be satisfied in both armchair
SWNTs and zigzag SWNTs.  For armchair SWNTs, though, the hopping integrals $%
\gamma _{y}\neq \gamma _{x}$ due to the wrapping and squashing.  The $%
\varphi _{y}$ can take any value near $\pi /3$ continuously as the restricted
condition is applied on $x$ direction.  So the condition $2\gamma _{y}\cos
\varphi _{y}=\gamma _{x}$ can always be satisfied.  Thus, a difference in the 
nearest-neighbor hopping integral along the axis and the circumference can not drive a
MST in armchair SWNTs.

However, for zigzag SWNTs, the system is restricted in $y$ direction, and $\varphi
_{y}$ can only take certain discrete values.  As a rough approximation, it can be assumed
that $\gamma _{y}=\gamma _{x}$ holds when a graphene sheet is wrapped into
a SWNT.  Thus, the condition $2\gamma _{y}\cos \varphi _{y}=\gamma _{x}$
changes to $\cos \varphi _{y}=1/2$, so $\varphi _{y}$ can only be $\pm \pi /3$.
This is the reason for the well-known result that zigzag ($n,0$) SWNTs are
metallic only when $n$ is a multiple of 3. (In fact, as mentioned above,
during the process in which a graphene sheet is wrapped into a SWNT, it is impossible to maintain $%
\gamma _{y}=\gamma _{x}$.  So even when $n$ is a multiple of 3,
there is a very narrow energy gap near the Fermi energy of zigzag ($n,0$%
) SWNTs, as shown in Figure 3a.)  For metallic zigzag SWNTs, when $\gamma
_{y}\neq \gamma _{x}$ due to the squashing, the condition $2\gamma _{y}\cos
\varphi _{y}=\gamma _{x}$ can not be satisfied through adjusting $\varphi
_{y} $, as $\varphi _{y}$ can only take certain discrete values.  Thus, a MST is
induced.

Now these results are demonstrated numerically by simulating the squashing of a zigzag (12,0) SWNT, as shown
in Figure 3.  To squash the tube, simulations are carried out in which two identical tips with a width of $d_{x}=5.0$
\AA\ are used to press the tube symmetrically about its center in the $\pm y$
direction, as shown in Figure 3b.  The tips are assumed to be rigid and to contact the nanotube through a hard-wall interaction.  The simulations, for both structural
optimization and calculation of electronic transport property, are performed
using a four-orbital tight-binding (TB) method, the details of which are reported in previous work.%
\cite{JQL} The typical conductance curve of a perfect zigzag (12,0) SWNT is
shown in Figure 3a.  As mentioned above, a very narrow gap is found near the
Fermi energy.  When the tube is squashed into an elliptical shape, as shown
in Figure 3b, a considerable gap can be found near the Fermi energy, indicating that
a MST takes place.  But to drive a MST in armchair SWNTs requires squashing of the nanotube
into a dumbbell shape so that the opposite walls interact.  So a MST can be achieved more easily by squashing
metallic zigzag SWNTs than by squashing armchair SWNTs.

\begin{figure}[t]
\begin{center}
\includegraphics[width=3.0in]{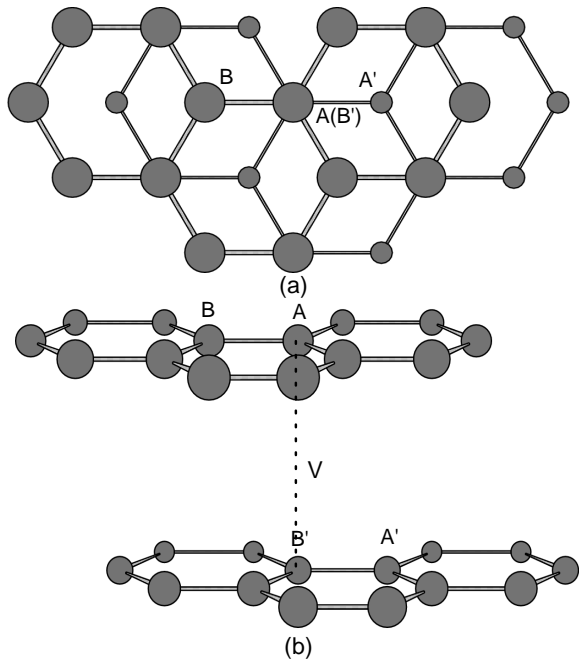}
\end{center}
\caption{The graphene bilayers with new interaction
introduced between the sublattics A and B'. (a) top view and (b) side view.}
\end{figure}

Last, it is emphasized that the effect of both of the two mechanisms is
to make the two original equivalent sublattices \textit{physically }%
distinguishable, as is the case in the one-dimensional lattice system.
In fact, the two sublattices in a graphene sheet or in a SWNT are originally
topologically distinguishable.  The \textit{physical} distinction means that
the two sublattices are `felt' differently by the conducting electrons,
which leads to a MST.  The first mechanism, the different onsite energies, clearly renders the two sublattices physically distinguishable.  And the other mechanism, the difference in nearest-neighbor hopping integral along the axis and the circumference, also makes the sublattices distinguishable, in the following way.  The condition $2\gamma _{y}\cos
\varphi _{y}=\gamma _{x}$ in a graphene sheet can be understood to enforce the
equality of the hopping integrals along the $x$ direction (the axis
direction of zigzag SWNTs).  When the condition can not be satisfied due to
the different between $\gamma _{x}$ and $\gamma _{y}$, the equality is
broken, which leads to the \textit{physical} distinction of the two
sublattices, as in the one-dimensional lattice system.  But for armchair
SWNTs, along the axis direction ($y$ direction), the equality of the hopping
integrals can always be satisfied.  Thus, the second mechanism can not
drive a MST in armchair SWNTs. \ As an aside, it might be concluded that the \textit{%
physical} distinction between the two sublattices is a general mechanism which
can drive a MST in any metallic SWNTs or even in the graphene layers.  But while it can drive a MST in a metallic SWNT, it can not drive a MST in graphene bilayers. For example, as
shown in Figure 4, a new interaction $V$ is introduced between the sublattices A
and B' of the graphene bilayers.  Clearly, in each layer, the two sublattices
are \textit{physically }distiguishable.  The energy dispersion relation
of the bilayer system can be expressed as:%
\[
\varepsilon -\varepsilon _{0}=\frac{\mp V\pm \sqrt{V^{2}+4\gamma ^{2}}}{2},
\]%
\[
\gamma ^{2}=\gamma _{0}^{2}\left( 1+4\cos ^{2}\varphi _{y}+4\cos \varphi
_{x}\cos \varphi _{y}\right) ,
\]%
where $\varepsilon _{0}$ is the onsite energy of the carbon atoms in the
graphene bilayers and $\gamma _{0}$ is the nearest-neighbor hopping integral.
In fact, $\varepsilon -\varepsilon _{0}=\gamma $\ is the energy dispersion
relation of a graphene monolayer.  A graphene monolayer is metallic, so $%
\gamma $\ can get the value $0$ with given value of $\varphi _{x}=\pm \pi $
and $\varphi _{y}=\pm \pi /3$.\  Then, at the same time, for the bilayer
system, $\varepsilon -\varepsilon _{0}=0$.\  This indicates that the system is
still metallic and that no MST is achieved.\  The results show the difference
between SWNTs and graphene bilayers. 

In summary, the two mechanisms which induce the MST in metallic SWNTs, i) a
difference in the onsite energies of the two sublattices, and ii) a
difference in the nearest-neighbor hopping integral along the axis
and the circumference , are theoretically analyzed through a simple tight-binding model.
The results prove that the MST can be achieved more easily in metallic zigzag
SWNTs than in armchair SWNTs; only the the first mechanism can lead to the MST
in armchair SWNTs, but both mechanisms can drive the MST in
metallic zigzag SWNTs.  Furthermore, the effects of the two
mechanisms are essentially the same: to make the two original equivalent sublattices 
\textit{physically }distinguishable.  This is the general mechanism that can
lead to a MST in any metallic SWNT and even in the graphene monolayer, but it
can not cause a MST in graphene bilayers.

The work is supported by the Ministry of Education of China, the National
High Technology Research and Development Program of China (Grant No.
2002AA311153).  H.T.J. acknowledges the support of NSF NER Grant DMR 02-10131.


\begin{thebibliography}{99}

\bibitem{SI} S. Iijima, Nature (London) \textbf{354}, 56 (1991).

\bibitem{JWM} J.W. Mintmire, B.I. Dunlap, and C.T. White, Phys. Rev. Lett. 
\textbf{68}, 631 (1992); N. Hamada, S.I. Sawada, and A. Oshiyama, Phys. Rev.
Lett. \textbf{68}, 1579 (1992).

\bibitem{TWT} T.W. Tombler, C. Zhou, L. Alexseyev, J. Kong, H. Dai, L. Liu,
C.S. Jayanthi, M. Tang, S. Wu, Nature (London) \textbf{405}, 769 (2000).

\bibitem{JQL} J.Q. Lu, J. Wu, W. Duan, F. Liu, B.F. Zhu, and B.L. Gu, Phys.
Rev. Lett. \textbf{90}, 156601 (2003); J.Q. Lu, J. Wu, W. Duan, B.L. Gu, 
Appl. Phys. Lett. \textbf{84}, 4203 (2004).

\bibitem{CJP} C.J. Park, Y.H. Kim, K.J. Chang, Phys. Rev. B \textbf{60},
10656 (1999).

\bibitem{PEL} P.E. Lammert, P. Zhang, V.H. Crespi, Phys. Rev. Lett. \textbf{%
84}, 2453 (2000).

\bibitem{EDM} E.D. Minot, Y. Yaish, V. Sazonova, J.Y. Park, M. Brink, P.L.
McEuen, Phys. Rev. Lett. \textbf{90}, 156401 (2003).

\bibitem{JC} J. Cao, Q. Wang, H. Dai, Phys. Rev. Lett. \textbf{90}, 157601
(2003).

\bibitem{MO} M. Ouyang, J.L. Huang, C.L. Cheung, C. M. Lieber, 
Science \textbf{292}, 702 (2001).

\bibitem{ARU} A. Rubio, J.L. Corkill, M.L. Cohen, Phys. Rev. B \textbf{49},
5081 (1994).

\end{thebibliography}
\end{document}